\documentclass[twocolumn,showpacs,preprintnumbers,amsmath,amssymb,nofootinbib]{revtex4-1}
\usepackage{here}
\usepackage{comment,braket}
\usepackage{epsf}
\usepackage{amsmath}
\usepackage{graphics}
\usepackage{amsfonts}
\usepackage{amssymb}
\usepackage{latexsym}
\usepackage{color}
\usepackage{ulem}
\input{colordvi.tex}
\usepackage{feynmp}

\usepackage{natbib}
\usepackage[pdftex]{graphicx}
\usepackage{bm}
\usepackage[pdftex]{hyperref}
\usepackage[svgnames]{xcolor}
\definecolor{phthaloblue}{rgb}{0.0, 0.06, 0.54}
\hypersetup{
    colorlinks=true,
    linkcolor=phthaloblue,
    citecolor=blue,
    filecolor=blue,
    urlcolor=phthaloblue,
    }

\newcommand{\beq}{\begin{eqnarray}} 
\newcommand{\eeq}{\end{eqnarray}}

\def\({\left(}
\def\){\right)}
\def\[{\left[}
\def\]{\right]}

\def\nn{\nonumber \\}

\def\lmk{\left(}
\def\rmk{\right)}
\def\lkk{\left[}
\def\rkk{\right]}

\newcommand{\eq}[1]{Eq.~(\ref{#1})}
\newcommand{\bel}[1] {\begin{equation}\label{#1}}
\newcommand{\beal}[1] {\begin{eqnarray}\label{#1}}
\newcommand{\be}{\begin{equation}}
\newcommand{\ee}{\end{equation}}
\newcommand{\bea}{\begin{array}} 
\newcommand{\eea}{\end{array}}

\begin{document}

\title{Tunneling wave function of the universe II: the backreaction problem}

\author{Alexander Vilenkin}
\author{Masaki Yamada}

\affiliation{Institute of Cosmology, Department of Physics and Astronomy, 
Tufts University, Medford, MA  02155, USA}

\date{\today}

\begin{abstract}
The tunneling wave function of the universe is calculated exactly for a de Sitter minisuperspace model with a massless conformally coupled scalar field, both by solving the Wheeler-DeWitt equation and by evaluating the Lorentzian path integral.  The same wave function is found in both approaches.  The back-reaction of quantum field fluctuations on the scale factor amounts to a constant renormalization of the vacuum energy density.  This is in contrast to the recent suggestion of Feldbrugge {\it et al.} that the back-reaction should diverge when the scale factor gets small, $a\to 0$.  Similar results are found for a massive scalar field in the limit of a large mass.  We also verified that the tunneling wave function can be expressed as a transition amplitude from a universe of vanishing size with the scalar field in the state of Euclidean vacuum, as it was suggested in our earlier work.
\end{abstract}

\maketitle

\section{Introduction}

In quantum cosmology the entire universe is treated quantum mechanically and is described by a wave function, rather than by a classical spacetime. The wave function $\Psi(g,\phi)$ is defined on the space of all 3-geometries ($g$) and matter field configurations ($\phi$), called superspace.  It can be found by solving the Wheeler-DeWitt (WDW) equation
\beq
{\cal H}\Psi=0,
\eeq
where ${\cal H}$ is the Hamiltonian operator.  Alternatively, the wave function can be expressed as a path integral,
\beq
\Psi(g,\phi) = \int^{(g,\phi)} {\cal D}g~{\cal D}\phi ~e^{iS},
\label{pathint}
\eeq
where $S$ is the action.

The choice of the boundary conditions for the WDW equation and of the class of paths included in the path integral has been a subject of ongoing debate. The most developed proposals in this regard are the no-boundary~\cite{Hartle:1983ai} and the tunneling~\cite{Vilenkin:1986cy,Vilenkin:1987kf,Vilenkin:1994rn} proposals.\footnote{For early work closely related to the tunneling proposal, see Refs.~\cite{Vilenkin:1982de,Linde:1983mx,Rubakov:1984bh,Vilenkin:1984wp,Zeldovich:1984vk}.}  The debate around these proposals has recently intensified \cite{Feldbrugge:2017kzv, Feldbrugge:2017fcc,Feldbrugge:2018gin,DiazDorronsoro:2017hti,DiazDorronsoro:2018wro,Vilenkin:2018dch}, spurred by the work of Feldbrugge {\it et al.}~\cite{Feldbrugge:2017kzv, Feldbrugge:2017fcc,Feldbrugge:2018gin}, who pointed out that the path integral in (\ref{pathint}) can be rigorously defined with the aid of the Picard-Lefschetz theory (at least in minisuperspace models, where the number of degrees of freedom is truncated to a finite number).  

Our focus in this paper will be on the tunneling wave function of the universe.  It was defined in Refs.~\cite{Vilenkin:1986cy,Vilenkin:1987kf} by specifying a boundary condition for the WDW equation.  Roughly, $\Psi$ is required to include only outgoing waves at the boundary of superspace, except for the part of the boundary corresponding to vanishing 3-geometries (see Refs.~\cite{Vilenkin:1986cy,Vilenkin:1987kf} for more details).  This is supplemented by the regularity condition, requiring that $\Psi$ remains finite everywhere, including the boundaries of superspace,
\beq
|\Psi(g,\phi)|<\infty.
\label{regularity}
\eeq
The resulting wave function can be interpreted as describing a universe originating at zero size, that is, from `nothing'. 

It was conjectured in Refs.~\cite{Vilenkin:1984wp,Vilenkin:1994rn} that the same wave function can be expressed as a path integral (\ref{pathint}) with the integration taken over (Lorentzian) histories interpolating between a vanishing 3-geometry and a given configuration $(g,\phi)$ in superspace.  In the simple de Sitter minisuperspace model describing a spherical universe with a positive vacuum energy density this expectation was confirmed in Ref.~\cite{Halliwell:1988ik} and more recently in \cite{Feldbrugge:2017kzv} using the Picard-Lefschetz method. However, the situation with extensions of the de Sitter model to perturbative superspace, including scalar field and/or gravitational wave perturbations, is still a matter of dispute.

The tunneling wave function in a perturbative WDW approach has been discussed in Refs.~\cite{Vilenkin:1987kf,Vachaspati:1988as}, with the conclusion that the modes of free scalar and gravitational fields are described by Gaussian wave functions corresponding to de Sitter invariant (Bunch-Davies) quantum states.  On the other hand, Feldbrugge {\it et al.}~\cite{Feldbrugge:2017fcc} argued that the path integral version of the tunneling proposal predicts a runaway instability: the probability of quantum fluctuations of the fields grows with their amplitude, so the conjecture of~\cite{Vilenkin:1984wp} does not hold.  Similar claims about instability of the tunneling proposal have also been made in the earlier literature~\cite{Halliwell:1989dy}. 

We have addressed this issue in our recent paper~\cite{Vilenkin:2018dch}, where we showed that quantum field fluctuations in the tunneling wave function (\ref{pathint}) are well behaved if the action $S$ is supplemented with a suitable boundary term.  Inclusion of this term is in fact necessary.  The regularity condition (\ref{regularity}) requires that the mode functions satisfy the Robin boundary condition at $a\to 0$, where $a$ is the radius of the universe, and the boundary term must be chosen so that the variational problem is well defined.  The path integral then gives the same wave function as the WDW approach in \cite{Vilenkin:1987kf,Vachaspati:1988as}.

A related issue is the behavior of the mode functions $\phi_n(a)$ at $a\to 0$.  The tunneling wave function has two branches in the classically forbidden (under-barrier) region: one branch growing with $a$ and the other decreasing with $a$.  On the growing branch our boundary conditions select the modes satisfying $\phi_n(0)=0$.\footnote{The same modes are selected by the Hartle-Hawking wave function, which does not include a decreasing branch.}  But on the decreasing branch the mode function grow without bound at $a\to 0$, and some authors suggested that this may cause serious problems.

In the path integral approach,  Feldbrugge {\it et al.}~\cite{Feldbrugge:2018gin} have argued that such behavior of the mode functions is unacceptable because it makes the mode action infinite.  However, we showed in \cite{Vilenkin:2018dch} that inclusion of the boundary term renders the action finite.  We emphasize that inclusion of this term is not a matter of choice: it is dictated by our boundary conditions.  Another concern raised in Ref.~\cite{Feldbrugge:2018gin} is that the unbounded growth of modes would cause an infinitely strong back-reaction on the geometry.  The perturbative expansion would then break down when one tries to go beyond the linear perturbation theory considered in~\cite{Vilenkin:2018dch}.  

Here, we are going to show that the back-reaction is actually well under control.  We first note that the problem, if it exists, should be present in the case of a massless conformally coupled field, where the mode functions exhibit the same behavior.  
Moreover, the same behavior of the mode functions is obtained in the WDW approach, so one would expect the same back-reaction problem to arise there as well.  An attractive feature of this model is that it allows an exact solution, so the back-reaction problem can be completely analyzed. 

In the next section we consider a de Sitter model with a massless conformal scalar field in the WDW approach and show that the field back-reaction amounts to the usual renormalization of the vacuum energy density.  We also consider a massive field in the perturbative superspace framework and reach the same conclusion regarding the back-reaction in the limit of a large mass.
In Section III we evaluate the Lorentzian path integral for both massless and massive models.  This yields the same results as the WDW approach.  
We also verify that the tunneling wave function can be expressed as a transition amplitude from a universe of vanishing size with the scalar field in the state of Euclidean vacuum, as it was suggested in~\cite{Vilenkin:2018dch}. 
Our results are summarized and discussed in Sec. IV.

\section{WDW approach}

\subsection{Perturbative superspace}

We consider a closed FRW universe,
\beq
ds^2 = a^2(\eta)\left(N^2 d\eta^2- d\Omega_3^2\right),
\eeq
with a conformally coupled scalar field $\phi$.  Here, $a(\eta)$ is the scale factor (radius of the universe), $\eta$ is the conformal time, and $N$ is the lapse parameter, which is set to be constant.
The action for this model is given by 
\beq
&&S=\int \sqrt{-{g^{(4)}}} \, d^4 x \lmk \frac{R}{2}- \rho_v \rmk + S_m +S_B ,
\label{action}
\\
&&S_m = \int \sqrt{-g^{(4)}} \, d^4 x \left[ - \frac{1}{2}(\nabla\phi)^2-\frac{1}{2}m^2\phi^2 -\frac{1}{12}R\phi^2\right]. \nn
\eeq
Here, $\rho_v$ is the vacuum energy density, $S_B$ is the boundary term, and we use Planck units with $\hbar=c=1$ and $8\pi G=1$.  The boundary term is unimportant in the WDW approach; it will be specified in the next section.  

We expand the field $\phi$ as
\beq
\phi(x,t) = 
\sum \phi_n(t)Q_n(x) 
= \frac{1}{a(t)} \sum \chi_n(t)Q_n(x),
\eeq
\beq
\int Q_n Q_{n'}^* d\Omega_3 = \delta_{n n'},
\eeq
where $Q_{nlm}(x)$ are suitably normalized spherical harmonics and we have suppressed the indices $l,m$ for brevity.

The wave function of the universe $\Psi(a,\{\phi_n\})$ satisfies the WDW equation
\beq
\left[\frac{1}{24\pi^2} \frac{\partial^2}{\partial a^2}-6\pi^2 V(a)+\sum_n n^2 {\cal H}_n\right] \Psi=0.
\label{WDW}
\eeq
Here, 
\beq
V(a)=a^2-H^2 a^4 ,
\label{Va}
\eeq
$H^2 = \rho_v /3$, the scalar field Hamiltonian for the $n$-th mode is
\beq
{\cal H}_n = -\frac{1}{2}\frac{\partial^2}{\partial\chi_n^2}+\frac{1}{2} \left(n^2+m^2a^2\right)\chi_n^2,
\eeq
and $n^2$ is the mode degeneracy factor.  We also disregard the ambiguity of ordering the factors $a$ and $\partial /\partial a$.  This is justified when $\rho_v \ll 1$ and the scale factor can be regarded as a semiclassical variable.\footnote{With a suitable choice of factor ordering the WDW equation for the scale factor can be solved exactly and the semiclassical approximation is not necessary \cite{Vilenkin:1987kf}.}
With the modes $\chi_n$ treated as small perturbations, a solution of Eq.~(\ref{WDW}) can be expressed as a superposition of terms of the form \cite{Halliwell:1984eu,Wada:1986uy, Vachaspati:1988as}
\beq
 \Psi (a, \chi_n) = A \exp \lkk - {12\pi^2} S(a) - \frac{1}{2} \sum_n R_n (a) \chi_n^2 \rkk,~~~~ 
 \label{wavefunction}
\eeq
where $A$ is a normalization constant.  Substituting this in (\ref{WDW}) and neglecting terms ${\cal O}(\chi_n^4)$, we obtain  
\beq
 \lmk \frac{dS}{da}\rmk^2 - V(a) - \frac{\hbar}{12\pi^2} S'' + \frac{\hbar}{12\pi^2} \sum_n n^2 R_n= 0,~~~~
 \label{EoM:Sa}
 \\
 \lmk \frac{dS}{da} \rmk \lmk \frac{dR_n}{da} \rmk - R_n^2 + \omega_n^2 (a) -\frac{\hbar}{24 \pi^2}{R_n}''= 0.~~~~
 \label{EoMSn}
\eeq
Here we explicitly wrote the Planck constant $\hbar =1$ only to indicate the subleading terms in the WKB expansion.\footnote{For a more detailed discussion of WKB expansion in the WDW equation, see e.g. \cite{Hong:2002yf}.} 
The terms proportional to $S''$ and $R_n''$ are responsible for WKB pre-factors, while the last term in (\ref{EoM:Sa}) accounts for the back-reaction of quantum field fluctuations on the dynamics of the scale factor $a$.  We shall first focus on the leading semiclassical behavior, neglecting terms proportional to $\hbar$.

In the classically forbidden range $(a<H^{-1})$ it will be convenient to introduce a Euclidean conformal time variable $\tau$ via
\beq
\frac{da}{d\tau}=\frac{dS}{da}= \pm \sqrt{V(a)}.
\label{dadtau}
\eeq
With $V(a)$ from (\ref{Va}), this has the solution 
\beq
a(\tau)=(H\cosh\tau)^{-1}
\label{atau}
\eeq
or 
\beq
e^\tau = \frac{1\pm \sqrt{1-H^2 a^2}}{Ha}.
\label{taua}
\eeq
The upper and lower signs in Eqs.~(\ref{dadtau}), (\ref{taua}) correspond to the decreasing and growing branches of the wave function respectively.  Note that for $a\to 0$ we have $\tau\to -\infty$ on the decreasing branch and $\tau\to +\infty$ on the growing branch.  More specifically, 
\beq
\frac{Ha}{2} \approx e^{\mp \tau} ~~~ (\tau\to \pm\infty).
\eeq

The tunneling boundary condition requires that only an outgoing branch of the wave function should be present in the classically allowed range.  The relative magnitude of the three branches can then be determined using the WKB connection formulas at the turning point $a=H^{-1}$.

\subsection{Mode functions}

Turning now to Eq.~(\ref{EoMSn}) for $R_n$, we rewrite it in the leading semiclassical order as
\beq
\frac{dR_n}{d\tau} - R_n^2 + \omega_n^2 (a) = 0.
\label{EOM-Rn}
\eeq
This is a Riccati equation; it can be reduced to a linear equation by the standard substitution
\beq
R_n(\tau)=-\frac{{\dot\nu}_n}{\nu_n},
\label{Rn}
\eeq
where dots stand for derivatives with respect to $\tau$ and the functions $\nu_n(\tau)$ satisfy the free field equation
\beq
{\ddot\nu}_n -\omega_n^2\nu_n = 0.
\label{nueq}
\eeq

The regularity condition (\ref{regularity}) requires that the functions $R_n(a)$ should satisfy ${\rm Re} \{R_n(a)\}>0$.  It has been shown in Ref.~\cite{Vilenkin:2018dch} that this condition is enforced, provided that the mode functions satisfy the Robin boundary condition
\beq
\frac{{d\nu}_n}{d\tau}=-n\nu_n
\label{Robin}
\eeq
at $\tau\to \pm\infty$.  This selects the solutions 
\beq
\nu_n(\tau) \propto \exp(-n\tau) ~~~ (\tau\to \pm\infty).
\label{blowup}
\eeq
As we noted, $\tau\to \pm\infty$ corresponds to $a\to 0$, with the upper and lower signs corresponding respectively to the growing and decreasing branches of the wave function.  We then find that $\nu_n \propto a^{\pm n}$ at $a\to 0$.  Hence on the growing branch our mode functions $\nu_n$ are regular at $a=0$, while on the decreasing branch they grow without bound.

Note, however, that it follows from Eq.(\ref{Rn}) that on both branches of the wave function we have $R_n(0)= n$, so the wave function $\Psi (a,\{\chi_n\})$ is non-singular at $a\to 0$.  Furthermore, the back-reaction terms $R_n$ in Eq.~(\ref{EoM:Sa}) are all regular and show no sign of an infinite back-reaction.  One may still be concerned that this is an artifact of perturbative superspace and that the back-reaction problem would arise in higher orders of perturbation theory in $\chi_n$.  We address this issue in the next section, where we discuss the exactly soluble case of a massless field.

\subsection{Massless field: an exact solution}

For a massless field, $m=0$, the solutions $\nu_n(\tau) \propto \exp(-n\tau)$ are exact and the mode functions exhibit the same divergent behavior at $\tau\to -\infty$.  In this case the WDW equation separates, and solutions can be found in the form
\beq
\Psi(a,\{\chi_n\})=\psi(a) \prod_n \psi_n(\chi_n).
\eeq
Here, $\psi_n(\chi_n)$ are eigenstates of ${\cal H}_n$ with eigenvalues $(p_n+\frac{1}{2})n$, where $p_n$ is an integer occupation number indicating the excitation level of the mode $n$.  The scale factor wave function $\psi(a)$ satisfies
\beq
\left[\frac{1}{24\pi^2}\frac{\partial^2}{\partial a^2}-6\pi^2 V(a)+\sum_n n^3 (p_n +\frac{1}{2})\right] \psi(a)=0,
\nn
\label{aeq}
\eeq 
where the last term represents the back-reaction of the scalar field modes on the scale factor.

Eq.~(\ref{aeq}) can be rewritten as
\beq
\left[\frac{1}{24\pi^2}\frac{\partial^2}{\partial a^2}-6\pi^2 a^2+2\pi^2 a^4\left(\rho_v + \Delta\rho_v +\rho_r  \right)\right] \psi(a)=0.
\nn
\label{aeq2}
\eeq 
Here, 
\beq
\rho_r(a) = \frac{1}{2\pi^2 a^4}\sum_n n^3 p_n
\eeq
is the energy density of scalar radiation, which is present if some of the occupation numbers $p_n$ are non-zero, and
\beq
\Delta\rho_v = \frac{1}{4\pi^2 a^4} \sum_n n^3
\label{Deltarhov}
\eeq
is the correction to the vacuum energy density due to the zero-point oscillations of the field modes.

The sum in Eq.~(\ref{Deltarhov}) is divergent.  It can be regularized by introducing a cutoff at a physical momentum $k_{\rm max} =\Lambda$, which corresponds to the wavenumber $n_{\rm max} = a\Lambda$.  Approximating the sum over $n$ by an integral over $k=n/a$, we have
\beq
\Delta\rho_v \approx \frac{1}{4\pi^2} \int_0^\Lambda k^3 dk=\frac{\Lambda^4}{16\pi^2},
\label{Deltarhov2}
\eeq
which is independent of $a$, as it should be.

If $\rho_r\neq 0$, it becomes the dominant term at small $a$, and the back-reaction becomes very significant.  In this case another classically allowed region appears near $a=0$, so the wave function does not describe tunneling from `nothing'.  Instead, it describes a universe originating at a singularity, then bouncing and recollapsing or alternatively tunneling to large values of $a$~\cite{Hong:2002yf}.  The tunneling boundary conditions require that $p_n=0$.  Then it follows from Eq.~(\ref{aeq2}) that there is no back-reaction effect, except for a constant renormalization of the vacuum energy density.

With $p_n=0$ the mode wave functions are given by
\beq
\psi_n \propto \exp\left(-\frac{n}{2}\chi_n^2\right).
\label{Gaussian}
\eeq
These wave functions decrease exponentially with $\chi_n$, so the fluctuations are well behaved.

\subsection{Massive field back-reaction}

\label{sec:massive}

Back-reaction of a massive quantum field can be analyzed in the limit of $m\gg H$.  In this case an approximate solution of Eq.~(\ref{EoMSn}) is 
\beq
R_n(a)\approx \omega_n(a).
\eeq
This approximation is accurate, provided that
\beq
S' R_n' \approx \frac{m^2 a \sqrt{V(a)}}{\omega_n}\ll \omega_n^2.
\label{adiabatic-condition}
\eeq
It is easily verified that this is always satisfied for $m\gg H$.

In this limit the back-reaction term in the WDW equation (\ref{EoMSn}) is given by 
\beq
\frac{1}{12\pi^2}\sum_n n^2 R_n &\approx&  \frac{1}{12\pi^2} \int d n n^2 \sqrt{n^2 + m^2 a^2} 
 \nn
 &=& \frac{a^4}{12\pi^2} \int_0^\Lambda d k k^2 \sqrt{k^2 + m^2}, 
\eeq
where we have defined a new variable $k=n/a$ and a UV cutoff scale $\Lambda$.
As before, this term gives a constant correction to the vacuum energy density,
\beq
 \Delta \rho_v = \frac{1}{4\pi^2} \int_0^\Lambda d k k^2 \sqrt{k^2 + m^2}.
 \label{Deltarho-massive}
\eeq

For $m\lesssim H$ the analysis is more complicated and we will not attempt it here.  We note also that our regularization method (a momentum cutoff) is rather crude and could miss subtle effects like trace anomaly.  We expect that such effects can be recovered using, for example, the Pauli-Villars regularization, but we shall not attempt to do that in this paper.

\section{Path integral approach}

We now consider the model of a conformally coupled field in the path integral approach, starting with the massless case.  The wave function is now given by
\beq
\Psi(a_1,\chi_{n1})=
\int_0^\infty d N \int {\cal D} a e^{i S_g (a, N)} 
\prod_n \int {\cal D}\chi_n e^{iS_n [\chi_n; N]}
\nn
\label{Psiaphi}
\eeq
where
\beq
S_g[a,N]=6\pi^2 \int_{\eta_0}^{\eta_1}\left[-\frac{{\dot a}^2}{N}+Na^2 \left(1-H^2a^2\right)\right] d\eta
~~~~
\label{Sg_initial}
\eeq
is the gravitational part of the action,
\beq
 S_n [ \chi_n; N] = \frac12 \int_{\eta_0}^{\eta_1} d \eta 
  \left(\frac{1}{N}{\dot\chi}_n^2 -N n^2 \chi_n^2 \right) 
  + S_{Bn}
  ~~~~
  \label{S_n}
\eeq
is the action for the $n$-th scalar field mode, and the boundary term 
\beq
 S_{Bn} = \frac{in}{4\pi^2 a^3} \int_{{\cal B}_0} \sqrt{-g^{(3)}} \, d^3 y \chi_n^2 = \frac{in}{2} \chi_{n}^2 (\eta_0)
\label{BT}
\eeq
has been added at the lower boundary ${\cal B}_0 : \eta=\eta_0$.  As we already mentioned, this term in the action is necessary to make the variational problem consistent with the Robin boundary condition (\ref{Robin}).
There is no boundary term at the upper boundary ($\eta=\eta_1$), because a Dirichlet boundary condition is imposed there.
As in the WDW formalism, there is no direct coupling between the variables $a$ and $\chi_n$, but both $S_g$ and $S_n$ depend on the lapse function $N$, and this opens the possibility of back-reaction.

\subsection{Semiclassical wave function}

We decompose the modes $\chi_n(\eta)$ into a classical part and a quantum fluctuation part: 
\beq
 \chi_n (\eta) = \bar{\chi}_n (\eta) + \xi_n (\eta).
\eeq
The classical part $\bar{\chi}_n(\eta)$ satisfies the classical equation of motion 
\beq
\frac{1}{N^2} \ddot{\bar{\chi}}_n + n^2 \bar{\chi}_n =0 
 \label{EoM2}
\eeq
with the boundary conditions
\beq
{\dot{\bar\chi}}_n(\eta_0)=inN \bar{\chi}_{n} (\eta_0), ~~~ {\bar\chi}_n(\eta_1)=\chi_{n1}. 
\label{Robin2}
\eeq
The solution is
\beq
{\bar\chi}_n(\eta)=\chi_{n1}e^{inN(\eta-\eta_1)}.
\label{chibar}
\eeq

The path integral over $\chi_n$ can be represented as a product $\psi_n=\psi_{nc} \psi_{nq}$, where
\beq
\psi_{nc} = e^{iS_{nc}} 
\eeq  
\beq
\psi_{nq} =  \int {\cal D} \xi_n e^{i {\tilde S}_n [ \xi_n; N] }.
\label{Psi_nq}
\eeq
Here,
\beq
S_{nc} &=& S_n[{\bar\chi}_n;N]
\nn
&=& \frac{i}{2N} {\bar \chi}_{n1} {\dot{\bar \chi}}_{n1}
 -  \frac{i}{2N} {\bar \chi}_{n0} {\dot{\bar \chi}}_{n0} + iS_{Bn}({\bar\chi}_n) 
\eeq  
is the classical action for the solution ${\bar\chi}_n(\eta)$ and 
\beq
 {\tilde S}_n[\xi; N] = \frac12 \int_{\eta_0}^{\eta_1} d \eta 
  \left(\frac{1}{N}{\dot\xi}_n^2 -N n^2 \xi_n^2 \right). 
  \label{S_xi}
\eeq

The last two terms in $S_{nc}$ cancel out and the classical contribution to the wave function for $\chi_n$ becomes
\beq
\psi_{nc} \propto  \exp\left(-\frac{1}{2}R_n\chi_{n1}^2\right),
\label{psinR}
\eeq
where 
\beq
R_n=-\frac{i}{N}\frac{{\dot\chi}_{n1}}{\chi_{n1}} = n.
\label{Rnchi}
\eeq
The path integral in $\psi_{nq}$ is independent of $\chi_n$, so the $\chi_n$ dependence of $\psi_n$ is
\beq
\psi_n \propto \exp\left(-\frac{n}{2}\chi_{n1}^2\right),
\label{psi_n}
\eeq
the same as in the WKB approach (\ref{Gaussian}).

\subsection{Massless field back-reaction}
\label{BR}

Evaluation of the remaining path integral over $\xi_n (\eta)$ is similar to the standard calculation of functional determinants, as e.g. in Ref.~\cite{Coleman:1985rnk}, except the standard calculation assumes Dirichlet boundary conditions $\xi_n(\eta_0)=\xi_n(\eta_1)=0$, while in our case the boundary conditions are
\beq
{\dot\xi}_n(\eta_0)=inN\xi_n(\eta_0), ~~~ \xi_n(\eta_1)=0.
\label{BC}
\eeq
The path integral can be reduced to Gaussian integrals by expanding $\xi_n (\eta)$ 
into an infinite series of complete orthonormal functions $f_p(\eta)$ ($p = 1,2,\dots$) that satisfy these boundary conditions. 
However, we can find such a set of functions only if $N$ is pure imaginary. 
In addition, ${\rm Im} N$ must be negative, so that we can perform the Gaussian integral. 
We set $N = - i \tilde{N}$ with $\tilde{N}$ being real and positive 
and calculate the path integral. After that we analytically continue the result as a function of $\tilde{N}$ ($= iN$).

We expand $\xi_n$ as 
\beq
 \xi_n (\eta) = \sum_{p = 1}^\infty c_p f_p (\eta), 
 \label{expansion}
\eeq
where $c_p$ are real constants and normalize the functions by 
\beq
 \int_{\eta_0}^{\eta_1} d \eta f_p (\eta) f_{p'} (\eta) = \delta_{p p'}. 
\eeq
The boundary conditions determine the form of the functions as 
\beq
 f_p (\eta) = A_p \sin \lkk k_p \lmk \eta - \eta_1 \rmk \rkk, 
 \label{f_p}
\eeq
where $A_p$ are normalization constants. 
The frequency $k_p$ is determined by 
\beq
 - \tan \lmk k_p \Delta \eta \rmk = \frac{k_p}{n \tilde{N}}, 
\eeq
and is labeled by an integer $p$, 
where $\Delta \eta \equiv ( \eta_1 - \eta_0)$. 
It satisfies 
\beq
 \frac{\pi}{\Delta \eta} \lmk p - \frac{1}{2} \rmk < k_p < \frac{\pi}{\Delta \eta} p. 
\eeq
Note that $k_{p+1} - k_p \approx \pi / \Delta \eta$ in the limit of 
$n \tilde{N} \gg 1/ \Delta \eta$, which is the case for $\eta_0 \to - \infty$. 

Expanding $\xi_n$ as in \eq{expansion}, we can rewrite \eq{Psi_nq} as 
\beq
 \psi_{nq} &\propto& \int \prod_p d c_p 
 \exp \lkk - \frac12 \sum_p \lmk \frac{1}{\tilde{N}} k_p^2 + n^2 \tilde{N} \rmk c_p^2 
 \rkk
 \nn
 &\propto& \exp \lkk - \frac12 \sum_p \ln \lmk \frac{1}{\tilde{N}} k_p^2 + n^2 \tilde{N} \rmk \rkk, 
\eeq
where we disregard the normalization constant. 
Noting that $\eta_0 \to - \infty$ as $a(\eta_0) \to 0$ for the classical solution, 
we take a limit of $n \tilde{N} \gg 1/ \Delta \eta$, 
which allows us to approximate the infinite sum as an integral as follows: 
\beq
 \psi_{nq} &\propto& \exp \lkk - \frac12 \frac{\Delta \eta}{\pi} \int_0^\infty d k \ln \lmk \frac{1}{\tilde{N}} k^2 + n^2 \tilde{N} \rmk \rkk 
 \nn
 &=& \exp \lkk - \frac12 n \tilde{N} \int d \eta \rkk. 
 \label{psi_nq result}
\eeq
In the last line we used $\Delta \eta = \int d \eta$. 

Since the result is an analytic function of $\tilde{N}$, 
we can analytically continue the resulting function to the whole complex plane of $\tilde{N}$. 
Rewriting $\tilde{N}$ as $i N$, 
we obtain 
\beq
 \psi_{nq} \propto \exp \lkk - i n N \int d \eta \rkk. 
 \label{det}
\eeq
Combining this with \eq{Sg_initial} 
and taking into account the degeneracy factor $n^2$ with summation over $n$, 
we find that the wave function (\ref{Psiaphi}) reduces to
\beq
\Psi(a_1,\chi_{n1})=\int_0^\infty d N \int {\cal D} a e^{i S (a, N)}, 
\label{PI}
\eeq
where $S(a,N)$ is given by Eq.~(\ref{Sg_initial}) with the replacement
\beq
H^2 = \frac{1}{3}\rho_v \to \frac{1}{3}\left(\rho_v+\Delta\rho_v\right)
\eeq
and $\Delta\rho_v$ given by (\ref{Deltarhov}).  Thus, as before, the effect of quantum fluctuations amounts to a constant renormalization of the vacuum energy density.  
The path-integral (\ref{PI}) can now be calculated using the Picard-Lefschetz theory as in the de Sitter minisuperspace model, like it was done in Refs.~\cite{Halliwell:1988ik,Feldbrugge:2017kzv,Brown:1990iv}.

\subsection{Massive field back-reaction}

In this section, we calculate the backreaction of a massive scalar field. 
We add $N a^2(\eta) m^2 \chi_n^2$ in the parenthesis for $S_n$ in \eq{S_n}. We shall first do the path integral over $\chi_n$ in Eq.~(\ref{Psiaphi}) treating $N$ as an undetermined parameter and $a(\eta)$ as an unspecified function, so that integrations over $N$ and $a(\eta)$ can be performed afterwards.

As before, we represent the field $\chi_n(\eta)$ as a sum of a classical solution and a quantum fluctuation.  The classical solution ${\bar\chi}_n(\eta)$ satisfies the equation
\beq
 \frac{1}{N^2} \frac{d^2\bar{\chi}_n}{d \eta^2} + \omega_n^2 (\eta) \bar{\chi}_n = 0, 
 \label{massivefieldeq}
 \eeq
 where
 \beq
 \omega_n^2(\eta) \equiv n^2 + a^2 (\eta) m^2. 
\eeq
The corresponding classical action $S_{nc}$ can be found using integration by parts and the classical field equation (\ref{massivefieldeq}), 
\beq
 &&i S_{nc} = - \frac12 R_n \chi_{n1}^2. 
\eeq
Here $R_n$ is defined by 
\beq
 &&R_n = - \frac{i}{N} \frac{\dot{\chi}_{n1}}{\chi_{n1}}, 
\eeq
and satisfies \eq{EOM-Rn} with the replacement $d\tau \to - i N d\eta$. 
Note that since we do not specify the function $a(\eta)$ in this calculation, 
$S_{nc}$ (or $R_n$) should be regarded as functionals in terms of $a(\eta)$. 
Now we shall use the WKB approximation, $d R_n / d (N \eta) \ll  R_n^2$. 
Then an approximate solution is given by $R_n \approx \omega_n(a_1)$. 
Since this is independent of the form of the function $a(\eta)$ in the range of $(\eta_0, \eta_1)$, 
the classical part of the massive field does not affect the equation of motion for $a(\eta)$.

The quantum correction comes from \eq{Psi_nq} with 
\beq
 {\tilde S}_n[\xi; N] = \frac12 \int_{\eta_0}^{\eta_1} d \eta 
 \left(\frac{1}{N}{\dot\xi}_n^2 -N n^2 \xi_n^2 - N m^2 a^2 \xi_n^2 \right). 
 \nn
 \label{S_xi}
\eeq
The boundary conditions are the same as in \eq{BC}. 
This path integral can be calculated in the limit of large $m$ by using an adiabatic expansion. 

First, we divide the domain of integration $\Delta\eta=\eta_1-\eta_0$ into $K$ small intervals $\epsilon=\Delta\eta/K$, 
\beq
 \int_{\eta_0}^{\eta_1} d \eta [...] = \sum_{j = 0}^{K-1} \int_{\eta_j}^{\eta_{j+1}} d \eta [...], 
 \label{divided-Int}
\eeq
where $\eta_j=\eta_0+j\epsilon$.
Then let us focus on the $j$-th interval. 
When $\epsilon$ is small enough, we can treat the scale factor as a constant. 
In this case, we can calculate the path integral 
in the same way as we did in Sec.~\ref{BR} with the replacement of $\Delta \eta \to \epsilon$ 
and $n \to \omega_n(\eta_j)$, where $\omega_n(\eta)= \sqrt{n^2+m^2 a^2(\eta)}$.  The result is
\beq
\exp\left[-iN\epsilon \omega_n(\eta_j)\right].
 \label{interval}
 \eeq

As we noted above \eq{psi_nq result}, 
this calculation requires that the condition
\beq
\tilde{N} \sqrt{n^2  + m^2 a^2 (\eta) } \gg 1/ \epsilon
\label{cond1}
\eeq
is satisfied. Furthermore, the assumption
that the variation of the scale factor can be neglected in the interval of ($\eta_j, \eta_{j+1}$) implies that
\beq
 \frac{1}{\omega_n} \frac{d\omega_n}{d \eta} \ll \frac{1}{\epsilon}. 
\label{cond2}
\eeq
We can choose $\epsilon$ satisfying both conditions (\ref{cond1}) and (\ref{cond2}), provided that
\beq
 \frac{1}{\tilde{N} \omega_n} \frac{d\omega_n}{d \eta} \ll \sqrt{n^2 + m^2 a^2 (\eta)}. 
\label{cond3}
\eeq
Anticipating that the back-reaction will only renormalize the vacuum energy, we can estimate the left-hand side 
of Eq.~(\ref{cond3}) using the known results for the de Sitter minisuperspace model.  
Then Eq.~(\ref{cond3}) is equivalent to the adiabatic condition \eq{adiabatic-condition} 
and is satisfied for $m \gg H$. 

Combining the contributions of different time intervals and of different $n$, we find
\beq
&&\prod_n \int {\cal D}\chi_n e^{iS_n [\chi_n; N]} 
\nn
&&\approx \prod_n \exp\left[-iN\epsilon \sum_j \omega_n(\eta_j)\right] 
\nn
 &&\approx
\exp\left[- i N \sum_n n^2 \int_{\eta_0}^{\eta_1} d \eta \sqrt{n^2 + m^2 a^2 (\eta) } \right].
 \label{interval}
\eeq
After replacing summation over $n$ by integration and introducing a cutoff, as in Section II.D, this reduces to
\beq
\exp\left[ -2\pi^2 i N \Delta\rho_v \int_{\eta_0}^{\eta_1} d \eta a^4 (\eta) \right], 
\label{massiveresult}
\eeq
where $\Delta\rho_v$ is given by Eq.~(\ref{Deltarho-massive}).  Substitution of (\ref{massiveresult}) into Eq.~(\ref{Psiaphi}) amounts to renormalizing the vacuum energy density in the gravitational part of the action, in complete agreement with the WDW analysis.

\subsection{Boundary term as initial wave function}

In the above calculation, 
the boundary term is added to the action as a fundamental law 
and the Robin boundary condition is imposed on the quantum variable $\chi_n$. 
On the other hand, as it was noted in \cite{Vilenkin:2018dch}, 
one can interpret the boundary term as an initial condition for the wave function 
without imposing the Robin boundary condition. 
In this subsection, we focus on the massless case for simplicity.

In this case the path integral for the $n$-th mode becomes
\beq
\psi_n(\chi_{n1},N) = \int {\cal D}\chi_n e^{iS_n [\chi_n; N]}\psi_{\rm ini}(\chi_{n0}),
\eeq
where 
\beq
\psi_{\rm ini}(\chi_{n0})=\exp\left(-\frac{n}{2}\chi_{n0}^2\right)
\eeq
can be thought of as the wave function of the Euclidean vacuum.
Here, the integration is performed with Dirichlet boundary conditions 
\beq
 \chi_n(\eta_1)=\chi_{n1},~~~ \chi_n(\eta_0)=\chi_{n0}
\label{Dbc} 
\eeq
and the integration measure includes an integral over $\chi_{n0}$.

The classical part of the path integral is then given by 
\beq
 \psi_{nc} 
 = \int d {\bar \chi}_{n0} \exp &&\lkk \frac{i}{2N} {\bar \chi}_{n1} \dot{\bar \chi}_{n1} 
 \right.
 \nn
 &&\left.
 - \frac{i}{2N} {\bar \chi}_{n0} \dot{\bar \chi}_{n0}  - \frac{n}{2} {\bar \chi}_{n0}^2 \rkk, 
\eeq
where ${\bar \chi}_n (\eta)$ satisfies the classical equation of motion \eq{EoM2} with the boundary conditions (\ref{Dbc}).
The solution is
\beq
{\bar\chi}_n(\eta) = \frac{
\chi_{n1} \sin \lmk n N (\eta - \eta_0) \rmk
- \chi_{n0} \sin \lmk n N (\eta- \eta_1) \rmk 
}{\sin \lmk n N \Delta \eta \rmk}.
\nn
\eeq
Using this solution, we can rewrite $\dot{\chi}_{n1}$ and $\dot{\chi}_{n0}$ in terms of $\chi_{n0}$ and $\chi_{n1}$. 
As a result, the exponent is given by 
\beq
 \frac{i n}{2} \cot \lmk n N \Delta \eta \rmk \chi_{n1}^2 
 - \frac{i n}{\sin \lmk n N \Delta \eta \rmk} \chi_{n1} \chi_{n0}
 \nn
 + \frac{n}{2} \lmk i \cot \lmk n N \Delta \eta \rmk - 1 \rmk \chi_{n0}^2. 
\eeq
Assuming ${\rm Re} [ i \cot ( n N \Delta \eta) - 1] >0$, 
we can perform the integral over $\chi_{n0}$. 
The result is given by 
\beq
 \psi_{nc} 
 \propto 
 \exp \lkk - \frac{n}{2} \chi_{n1}^2 \rkk, 
\eeq
which is the same as \eq{psinR}. 
Although we assumed ${\rm Re} [ i \cot ( n N \Delta \eta) - 1] >0$, 
we expect that the result can be used for arbitrary $N$ 
by analytic continuation.

The boundary condition for the quantum fluctuation part $\psi_{nq}$ 
is also the Dirichlet boundary condition at the final and initial surfaces. 
The calculation of the quantum fluctuation part is similar to that in the previous section, except that $k_m$ is determined by 
\beq
 k_m = \frac{\pi m}{\Delta \eta}. 
\eeq
The result in the limit of $n \tilde{N} \gg 1/ \Delta \eta$ 
is the same as \eq{det}.

\section{Summary and discussion}

We discussed the tunneling wave function of the universe in de Sitter minisuperspace with a conformally coupled massless scalar field using both the WDW and path integral approaches.  We found by an exact calculation (i) that the two approaches give the same wave function and (ii) that the back-reaction of quantum field fluctuations on the scale factor amounts to a constant renormalization of the vacuum energy density $\rho_v$.  We also verified that the tunneling wave function can be expressed as a transition amplitude from a universe of vanishing size with the scalar field in the state of Euclidean vacuum, as it was suggested in \cite{Vilenkin:2018dch}.
Furthermore, we considered a massive conformally coupled field in the limit of large mass, $m\gg H$, and found that once again the back-reaction gives only a constant renormalization of $\rho_v$.  We expect the same conclusions to hold for arbitrary values of $m$, but the analysis in the general case would require more sophisticated regularization methods (e.g., Pauli-Villars), and we leave it for future work.

We now comment on why the divergence of mode functions at $a\to 0$ in the tunneling wave function does not result in infinite back-reaction, as it was expected by Feldbrugge {\it et al.} in Ref.~\cite{Feldbrugge:2018gin}.  These authors assumed that the effect of back-reaction can be accounted for simply by adding the classical energy-momentum tensor of the modes to the right-hand side of classical Friedmann equations.  This, however, does not appear to be the case.

The mode functions $\nu_n(\eta)$ are related to the wave function (\ref{wavefunction}) by Eq.~(\ref{Rn}).  In the classically allowed range ($a>H^{-1}$) these are the "negative energy" mode functions \cite{Wada:1986uy,Laflamme:1987mx}.  These mode functions are complex, and when substituted in the energy-momentum tensor for a real scalar field, they would give a complex $T_{\mu\nu}$.  Back-reaction of quantum fields on the metric has been extensively studied by calculating the expectation value $\langle T_{\mu\nu}\rangle$ in a classical spacetime (for a review see \cite{Birrell:1982ix}).  The contribution of a given mode $\nu_n$ to the expectation value $\langle T_{00}\rangle$ is given by
\beq
\frac{1}{4\pi^2 a^4}\left({\dot\nu}_n^* {\dot\nu}_n+n^2 \nu_n^* \nu_n\right).
\label{energy}
\eeq
With $\nu_n(t)=(2n)^{-1/2}\exp(in\eta)$, this gives 
\beq
\frac{n}{4\pi^2 a^4}, 
\label{energy2}
\eeq
which is real and agrees with Eq.~(\ref{Deltarhov}).
One can expect that the corresponding contributions on the two branches of the wave function in the classically forbidden range can be obtained from (\ref{energy}) by analytic continuation $\eta\to\pm i\tau$.  This gives the same result (\ref{energy2}) and no divergence.

We note also that even though the mode functions diverge at $a\to 0$, the functions $R_n(a)$ are finite, so the wave function of the universe (\ref{wavefunction}) is well behaved.  Furthermore, the functions $R_n(a)$ describe the effect of back-reaction in the WDW equation (\ref{EoM:Sa}); hence this effect is clearly finite, at least in the WDW approach.

Another objection that has been raised against the path integral form of the tunneling wave function is that it gives a Green's function (propagator) rather than a solution of the WDW equation \cite{DiazDorronsoro:2017hti,Halliwell:2018ejl}.  This, however, is not a valid distinction in the present case.  The delta function in the propagator equation is $\delta(a)$, so its support is at the boundary of superspace ($a=0$) and thus the propagator satisfies the WDW equation everywhere in superspace.  This is supported by our result that the path integral version of the tunneling wave function coincides with the WDW version.  

We finally comment on the most recent version of the no boundary wave function \cite{Halliwell:2018ejl}.  The original proposal \cite{Hartle:1983ai} was based on the Euclidean path integral, but it was soon realized that as it stands this integral is divergent, because the gravitational part of the Euclidean action is unbounded from below.  Attempts have been made to fix the problem by extending the path integral to complex metrics \cite{Halliwell:1988ik,Halliwell:1989dy}.  However, the space of complex metrics is very large and no obvious choice of the set of complex histories in the path integral suggests itself as the preferred one.  The no-boundary approach was recently applied to several minisuperspace models by Dorronsoro {\it et al.} \cite{DiazDorronsoro:2017hti,DiazDorronsoro:2018wro}, where they used different lapse integration contours in the complex plane for different models. (An extensive list of references to earlier literature can also be found in these papers.)  This analysis made it obvious that the no boundary proposal is incomplete without a choice of a complex integration contour in the path integral.  Some general requirements to this contour have been given in Ref.~\cite{Halliwell:1989dy}, but it is not clear that they can always be satisfied or what contour should be used in models admitting a number of choices that satisfy the requirements.

Most recently, Halliwell, Hartle and Hertog \cite{Halliwell:2018ejl} proposed yet another version of the no boundary wave function, apparently in an attempt to address the difficulties indicated above.  They suggest that the semiclassical wave function of the universe has the form
\beq
\Psi(g,\phi) \approx \sum_i d_i \exp\left( -S_i(g,\phi)\right).
\label{HHH}
\eeq
Here, $S_i(g,\phi)$ is the Euclidean action evaluated for a regular (generally complex) solution of Einstein's equations on a four-disk (a saddle point) with boundary conditions $(g,\phi)$ on its boundary.  The index $i$ labels different saddle points.  No attempt has yet been made to extend this wave function proposal beyond the semiclassical level. 

In this proposal there is no path integration, so one does not have to choose between different integration contours, but the choice of the coefficients $d_i$ appears to be arbitrary.  Halliwell {\it et al.} suggest that saddle points predicting unbounded quantum fluctuations should be excluded (assigned $d_i =0$) and that saddle points with actions $S_i$ and $S_i^*$ should contribute with equal weight.  But this still allows much room for different choices, especially if the model admits a large number of saddle points.  
In our view, the basic criticism against this approach still remains: it does not specify the wave function uniquely.  But the wave function of the universe should be unique, since if it is not, then what determines the choice between different alternatives?

\begin{acknowledgments}
This work was supported in part by the National Science Foundation.
\end{acknowledgments}

\bibliography{reference}

\end{document}